# Atomically Thin Amorphous Carbon with an Ultralow Dielectric Constant

Chee-Tat Toh[1,2,3,†], Artem K. Grebenko[2,3†], Ugur Karadeniz[1], Usha Bhat[1], Ya He[1], Hongji Zhang[1,2], Denis V. Vyalikh[4,5], Anna Makarova[6], Alexander Fedorov[7,8,9], Alena A. Alekseeva[2,3], Kostya Iakoubovskii[3], Lu Shi[1], Andrei Starkov[1,2], Chuan Chu Tee[1], Lucas M. Sassi[3], Michel Bosman[1,10], Naoto Kamiuchi[11], Yuta Sato[12], Kazutomo Suenaga[11], Barbaros Özyilmaz[1,2,3,13]

[1]Department of Materials Science and Engineering, National University of Singapore, Singapore

[2]Centre for Advanced 2D materials, National University of Singapore, Singapore

[3]Department of Physics, National University of Singapore, Singapore

[4]Donostia International Physics Center (DIPC) Donostia-San, Sebastián, Spain

[5]IKERBASQUE Basque Foundation for Science, Bilbao, Spain

[6]Physikalische Chemie, Institut für Chemie und Biochemie, Freie Universität Berlin Arnimallee 22, Berlin, Germany

[7]Leibniz Institute for Solid State and Materials Research, Dresden, Germany

[8]Helmholtz-Zentrum Berlin für Materialien und Energie, Berlin, Germany

[9]Joint Laboratory "Functional Quantum Materials" at BESSY II, Berlin, Germany

[10]Institute of Materials Research and Engineering, Agency for Science, Technology and Research (A*STAR), Singapore

[11]Sanken, Osaka University, Ibaraki, Japan

[12]Nanomaterials Research Institute, National Institute of Advanced Industrial Science and Technology (AIST), Tsukuba, Japan

[13]Institute for Functional Intelligent Materials, National University of Singapore, Singapore

[†]These authors contributed equally to this work.

***Two-dimensional (2D) materials exhibit excellent properties at monolayer thickness and are viable replacements for various microelectronic components as scaling gradually approaches the atomic limit[1–3]. Despite significant advancements in the ongoing 2D revolution of integrated circuits[4–7], one crucial building block, namely a 2D ultralow-k[8] (ULK) dielectric, remains unreported. The challenge lies in achieving a dielectric constant (k) less than 3, as traditional low-k dielectrics are inherently unstable at the 2D limit due to their amorphous[9] or porous[10] nature. The realisation of ultrathin dielectrics with low-k***

***is also needed to address current bottlenecks in integrated circuits scaling[11,12]. Specifically, low-k materials are necessary to minimise parasitic capacitances[13] as the distance between conductive elements shrinks below 10 nm[14,15]. Moreover, advanced architectures like gate-all-around field effect transistors (GAA FET) require even lower dielectric constants (k<2) at sub-3nm thickness[15]. Here, we show that layer-by-layer grown multilayer amorphous carbon (ML-AC), as thin as 0.8 nm, is a mechanically robust 2D ULK dielectric with k of 1.35 and dielectric strength of 28-31 MV cm$^{-1}$. The lack of any long-range order, its intrinsic 2D nature, sp$^2$ carbon character and low density are all essential for minimising dielectric permittivity. Moreover, ML-AC overcomes the vulnerability of existing dielectrics to ion diffusion degradation with a record metal ion diffusion time to failure (TTF) of 10$^{10}$ s for even a single layer. Therefore, otherwise necessary additional layers occupying up to 3 nm can be eliminated, which is especially significant as metal line widths approach 10 nm. Combined with its low-temperature, direct and conformal growth even on a dielectric, these critical features enable substantial improvements in silicon-based semiconductor electronics and ensure compatibility with future 2D electronics.***

The 2D materials toolbox[2,3] offers a comprehensive range of properties, making them highly versatile for various device functionalities. 2D crystals exhibit diverse properties[16–18] such as insulating, metallic, semiconducting, superconducting, ferromagnetic, and ferroelectric. However, one critical component for advanced electronics is still missing: an atomically thin ultralow-k dielectric[19]. The search for a 2D ULK dielectric among crystalline systems is fundamentally limited by the lattice order, which favours dipoles to align. On the other hand, the introduction of porosity and disorder in 2D remains structurally challenging. Current solutions are vulnerable when used at thickness below 10 nm, from either losing their ULK properties or overall mechanical integrity during fabrication processes[20]. Furthermore, porous structures are not only susceptible to metal ion diffusion, but they also hinder the formation of highly crystalline metal lines directly on them.

2D ULK dielectric would have an immediate impact for the silicon-based 3D semiconductor industry, irrespective of the ongoing 2D revolution[1,4–6]. The International Roadmap for Devices and Systems (IRDS) has identified several scientific and technological challenges that must be overcome to increase the performance and scale down of existing integrated circuits, and to develop future platforms. First, the next generation of metal interconnects, with narrower inter-metal dielectric, requires relative dielectric permittivity of 2 or less to minimise parasitic capacitance between conductive elements[15]. Second, the realisation of 3D transistor architectures, such as GAA transistors, requires shrinking ULK materials to less than 3 nm in thickness[15]. Ultimately, a 2D ULK dielectric functioning down to monolayer thickness is highly desirable.

Dielectric permittivity is primarily influenced by two factors: material density and polarizability[21]. Previous research has focused on decreasing k by introducing low density materials or by creating a nanoporous structure. Notably, covalent organic frameworks with molecular porosity have reached values as low as 1.6[10]. However, scaling down nanoporous films below 20 nm thickness is challenging due to their reduced hardness. Alternatively, amorphous materials can also reduce dielectric permittivity without the need for pores, by restricting dipole alignment through disorder and thus decreasing polarizability. For thinner films, 3D amorphous boron nitride (a-BN) achieved k as low as 1.78 at 3 nm thickness[9]. However, it is thickness dependent[,22,23], and thus scaling down to the 2D limit while maintaining its ultralow-k value remains to be seen.

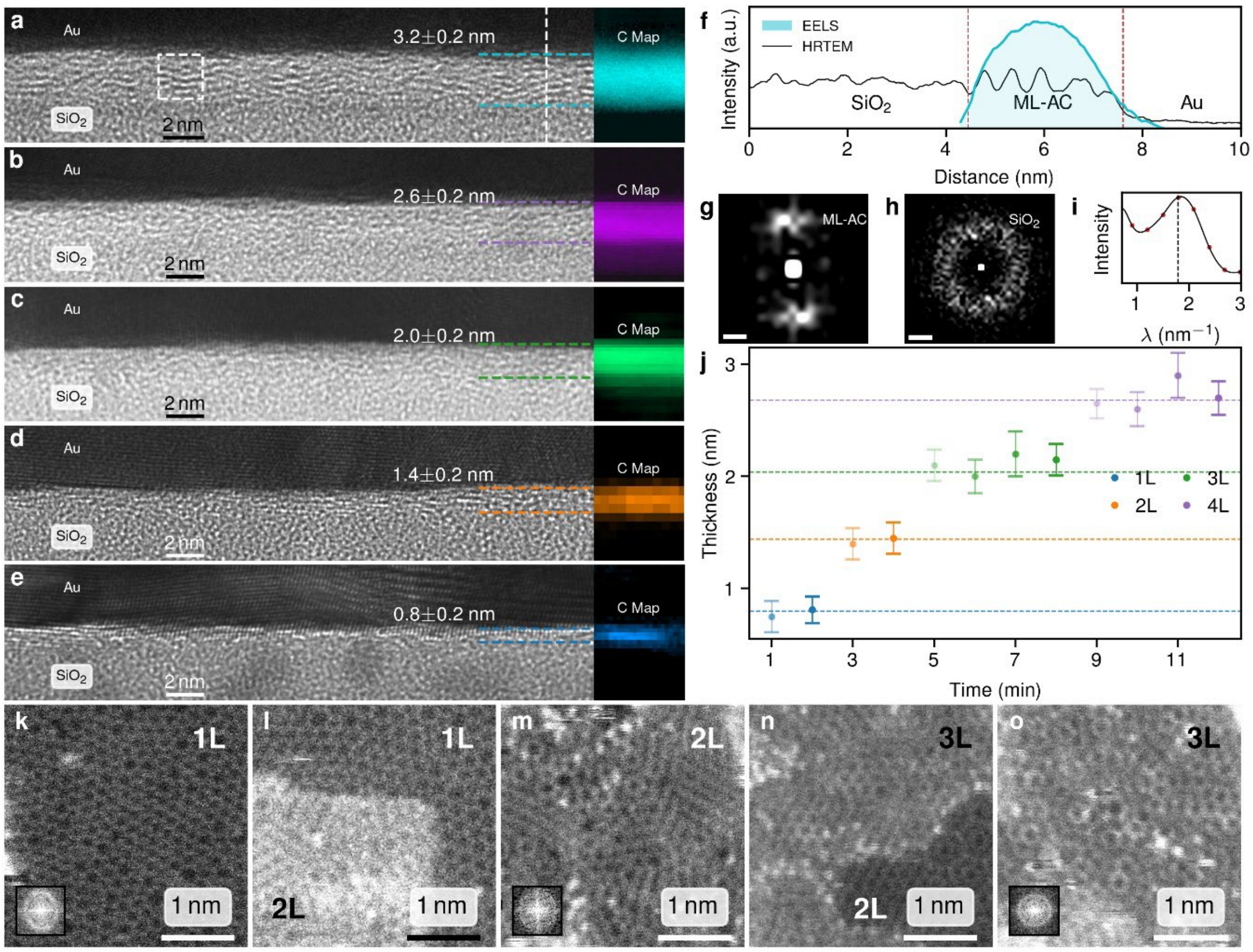

**Figure 1. Layered structure of ML-AC. a-e**, Cross-sectional HRTEM images of ML-AC on $SiO_2$/Si wafer with corresponding Carbon K-edge EELS elemental map for 5- ,4- ,3- , 2- and 1- layer thickness respectively. **f,** HRTEM intensity line profile averaged over 10 pixels from **a** (dashed white line) and EELS intensity line profile extracted from the corresponding Carbon elemental map. FFT image of **g,** from boxed region in **a,** and **h,** $SiO_2$ region. **i,** FFT spectra of the HRTEM profile shown in **f**, with maximum at 1.6 nm$^{-1}$, corresponding to 0.6 nm in real space. **j,** Thickness of directly grown ML-AC on $SiO_2$ as a function of growth duration measured by AFM. Dashed lines represent discrete intervals of 0.6 nm, starting from 0.8 nm. **k-o**, STEM atomic resolution images of ML-AC grown for 2, 3, 4, 6, and 8 min, respectively, showing complete (**k, m, o**) and incomplete (**l, n**) top layer. Inset: FFT image. The completely disordered structure is shown for **k**, 1 layer with random carbon rings, **m**, 2 layers and **o**, 3 layers with local moiré interference lines and rings.

Here we report layer-by-layer growth of multilayer amorphous carbon (ML-AC) that is composed of stacked monolayer amorphous carbon (MAC)[24,25], as a solution for ULK dielectrics. We achieve ultralow-k by significantly reducing the total polarizability of the material[22]. While orientational and electronic contributions to the polarizability are suppressed by structural disorder, ionic contribution is minimised by using mono-elemental carbon since carbon-carbon bonds are nonpolar[26]. In addition, ML-AC has lower k values because its sp$^2$ bonds consist of smaller hybridised orbitals than the sp$^3$ bonds dominant in 3D amorphous carbon[27], which makes it less susceptible to polarisation[28]. Furthermore, ML-AC has a low density without the introduction of pores, due to out-of-plane buckling intrinsic to its 2D structure, which increases the interlayer spacing from 0.34 nm for its crystalline graphene counterpart to ~0.6 nm. With approximately the same number of carbon atoms within the 2D structure, this effectively reduces the density from 2.3 g $cm^{-3}$ to the range of 1.0-1.3 g $cm^{-3}$, allowing ML-AC to obtain a lower k value. In addition, it maintains structural integrity and desirable dielectric properties down to a monolayer.

Equally important, we present a significant advancement in the growth of ML-AC, demonstrating its conformal growth at temperatures below 300 °C on arbitrary surfaces using chemical vapour deposition (CVD). We have successfully transitioned from self-limited, single-layer growth on metallic substrates to surface-independent, multilayer growth. This was achieved by substituting methane with acetylene as the carbon precursor and implementing a remote inductively coupled plasma. These modifications shifted the carbon radical species from primarily C and C-H to predominantly $C_2$ radicals, thereby increasing the total radical concentration. We examined ML-AC, which was directly grown on $SiO_2$, using cross-sectional high resolution transmission electron microscopy (HRTEM) complemented by elemental mapping through electron energy loss spectroscopy (EELS) (**Fig. 1a-e**). The EELS spectra for all film thicknesses showed the prominent $sp^2$ shoulder at 284.4 eV with a similar intensity ratio to the main carbon peak[24] (**Extended Data Fig. 1**). Samples exhibited thickness of 3.2 nm, 2.7 nm, 2.1 nm, 1.4 nm and 0.8 nm and high uniformity with thickness variation of 0.2 nm. For all thicknesses, the layers in ML-AC were resolved, with the 3.2 nm thickness providing the best resolution for further analysis (**Fig. 1f** and **Extended Data Fig. 1**). The FFT image of ML-AC (**Fig. 1h**) shows the 2-fold symmetry characteristic of a layered system, contrasting with the amorphous halo observed for $SiO_2$ (**Fig. 1i**). In addition, the interlayer spacing of ML-AC was determined to be 0.6 nm across all thicknesses, as obtained from Fourier transform analysis (**Fig. 1g**).

We employed a combination of atomic force microscopy (AFM) thickness measurements and top-view atomic resolution scanning TEM (STEM) imaging to obtain additional evidence for the layered structure ML-AC. We prepared a series of ML-AC samples with growth times increasing in 1 min steps. The thickness of more than 40 ML-AC samples were measured by AFM (**Extended Data Fig. 2**), revealing discrete thickness values of 0.8 nm, 1.4 nm, 2.0 nm and 2.6 nm (**Fig. 1j**). This well-defined, step-like increase in thickness of ~0.6 nm is in good agreement with the cross-sectional HRTEM results and corresponds to the known thickness of MAC[24]. This observation is further supported by atomic resolution STEM, which reveals the growth of continuous 1-, 2-, and 3-layer ML-AC (**Fig. 1k-o**). The 1-layer ML-AC features distorted nanocrystallites embedded within a continuous random network, resembling structures previously reported in MAC studies.[24,25] For all films, the Fourier transform (insets of **Fig. 1k, m, o**) displayed a broad, continuous halo instead of discrete diffraction spots, confirming the absence of long-range order in ML-AC. The correlation of STEM and AFM results suggests the formation of complete 2D layers of MAC, stacked one on top of another. In contrast, a 3D amorphous film structure would typically exhibit intermediate thickness due to the growth of 3D island clusters.

All STEM images displayed a fully connected, in-plane random carbon ring structure. The structure of 1-layer ML-AC was identical to that reported earlier for MAC grown on copper substrates. Interestingly, 2- and 3-layer ML-AC exhibited local moiré interference, with features similar to layer-by-layer transferred MAC (**Extended Data Fig. 3**). The distinct moiré lines and rings observed can only arise if the overlapping layers consisted of locally strained carbon rings of varying size or orientation[29]. This is representative across more than 10 samples from different growths, and additional STEM data for the structure of 2- and 3-layer ML-AC has been included in **Extended Data Fig. 3** respectively.

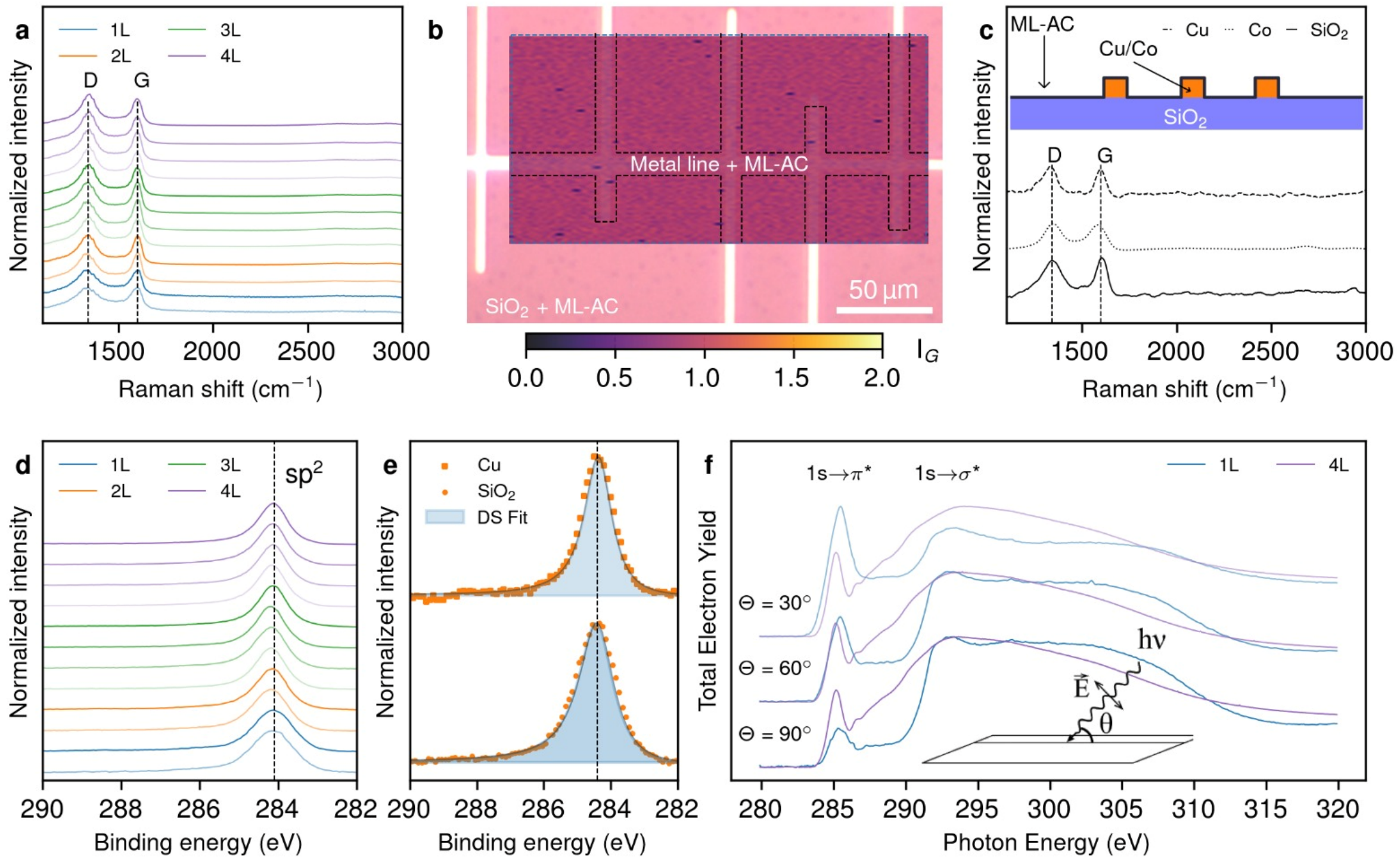

**Figure 2**. **Uniform growth of ML-AC. a,** Raman spectra of ML-AC on $SiO_2$ for different growth durations

(matching **Fig. 1j**). **b**, Optical image of lithographically defined array of Co lines covered with ML-AC, with an overlay showing Raman map of G-band intensity across the surface of $SiO_2$ and Co lines. Schematics of structure is shown in the inset of **c**. **c**, Individual Raman spectra of ML-AC directly grown on $SiO_2$ (solid line), Cu (dashed line) and Co (dotted line). **d,** Core-level XPS C 1s spectra of ML-AC on $SiO_2$ for different growth durations (matching **Fig. 1j**). **e,** Core-level XPS C 1s spectra of 2-layer ML-AC grown on Cu (squares) and $SiO_2$ (circles). Doniach-Sunjic function fit of the $sp^2$ peak at 284.4 eV is shaded in blue **f,** NEXAFS C-K edge spectrum showing weak angular dependence of the 1s-π* and 1s-σ* transition for 1- (blue) and 4-layer (purple) ML-AC. Inset: Measurement geometry with incident angle.

The ML-AC samples were further characterised for layer-dependent uniformity. The Raman spectra (**Fig. 2a**) of ML-AC are independent of thickness, exhibiting $I_D/I_G$ ratios close to the expected value of 0.85 for MAC[24]. They also displayed similar shapes and positions of the D and G peaks, as expected for a discrete layer system. When ML-AC was grown over a lithographically defined array of Cu and Co electrodes, the Raman map of the G-band (**Fig. 2b)** and the individual spectra (**Fig. 2c)** were found to be comparable on both the metals and $SiO_2$. 2-layer ML-AC was grown over large areas (**Fig. 1d**), and AFM measurements revealed a consistent surface roughness of 210 pm (**Extended Data Fig. 4**). This is indistinguishable from the underlying $SiO_2$ substrate and indicates that ML-AC conforms to $SiO_2$ during its growth process[31].

X-ray photoelectron spectroscopy (XPS) was used to determine the elemental composition and bond character of ML-AC. The C 1s core level spectra, captured from ML-AC on $SiO_2$ across various growth durations (**Fig. 2d**), and from 2-layer MAC grown on $SiO_2$ and Cu (**Fig. 2e**), were fitted with Doniach-Sunjic function, typically used for $sp^2$ carbon contribution[32]. The exact fit (line position at 284.4 eV) to the data indicates that ML-AC consists of mainly $sp^2$ carbon bonds with other contributions being negligible (<3%). The different layers of ML-AC reveal identical XPS spectra, confirming that the structure of ML-AC closely resembles a stack of MAC layers without interlayer bonds. In addition, we measured core level C 1s spectra of ML-AC from other growth substrates, such as Cu and Co, as well as core level spectra of the substrate materials. These measurements show the absence of substrate-carbon bond formation (**Extended Data Fig. 4**).

The amorphous $sp^2$ structure of ML-AC was further investigated using angle-dependent near-edge X-ray absorption fine structure (NEXAFS). The C K-edge NEXAFS spectra of directly grown ML-AC on $SiO_2$ were collected at various incident angles between the beam and the sample surface (**Fig. 2f**). Graphene typically exhibits a characteristic sharp 1s-π* transition at 285.5 eV, which has the strongest intensities at grazing angles and vanishes at 90° [30]. In comparison, the 1s-π* transition of ML-AC does not entirely disappear at 90°, which we attribute to the in-plane disordered structure and out-of-plane buckling of the $sp^2$ bonds. In addition, the spectra of 1- and 4-layer ML-AC both display broad peaks for both $\sigma$* and π* states, indicating the presence of a similarly disordered layered structure.

Therefore, we conclude that ML-AC consists of individual layers of MAC, each with a similar 2D amorphous carbon structure. The growth of ML-AC on metals like Cu and Co, as well as dielectrics like $SiO_2$, suggests a largely substrate-independent growth mechanism.

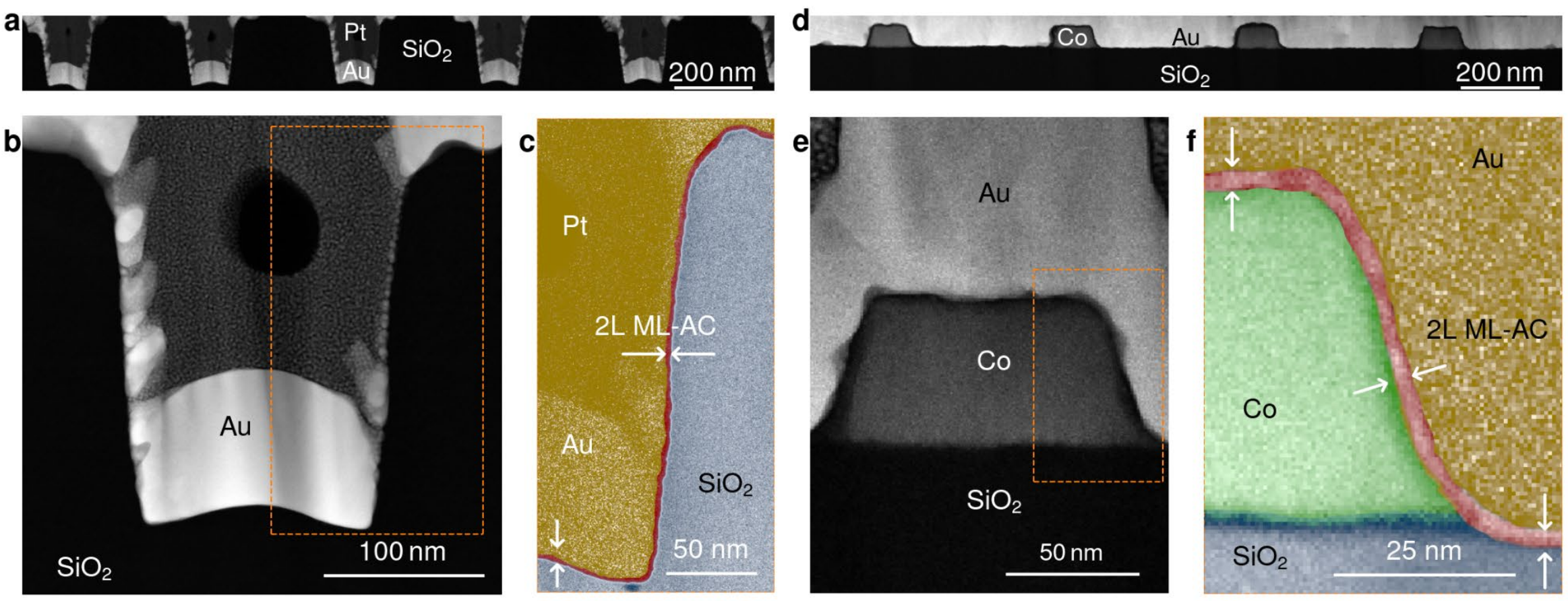

**Figure 3. Conformal coating by ML-AC. a, b,** STEM and **c**, EELS map corresponding to the box area in **b**, shows conformal coating of silicon dioxide trenches with 2-layer thick ML-AC. **d, e,** STEM and **f**, EELS map corresponding to the box area in **e**, shows conformal coating of Co lines and $SiO_2$ substrate with 2-layer thick ML-AC.

Next, we evaluate the potential of ML-AC for use in metal interconnects. The multiple interconnect layers, which provide conductive pathways down to the bottom transistor level (**Extended Data Fig. 5**), currently require diffusion barriers and liner layers on the sides, capping layers on the top, and a surrounding low-k dielectric. These layers occupy a significant volume of the interconnect, and the ideal solution would be to replace all of these with a single, atomically thin material that can be conformally grown.

We demonstrated the conformal growth of ML-AC by evaluating its encapsulation around structures that replicate the complexity of interconnects. Cross-sectional STEM coupled with EELS revealed that a 2-layer ML-AC (1.4 nm) was uniformly grown on the bottom and sidewall surfaces in silicon dioxide trenches that are 100 nm wide and 250 nm deep (**Fig. 3a-c**). This suggests that the growth of ML-AC is unaffected by topological features or confinement. To test the limits of conformality, a 2-layer ML-AC was also grown on Co lines with a width of 100 nm, height of 60 nm and pitch of 400 nm on a $SiO_2$/Si substrate. Despite the challenges posed by the abrupt transition at the metal and dielectric interface, we observed a uniform and conformal ML-AC coating across the Co lines and $SiO_2$ interface using cross-sectional STEM with EELS (**Fig. 3d-f**).

To demonstrate the absence of microscopic cracks and other defects, particularly at the sidewalls, we take advantage of the high sensitivity of Co and Cu to oxygen, water and etchants. Two sets of Co electrodes, with and without an ML-AC layer, were exposed to air for 72 hours. Subsequent AFM Co lines significantly expanded ~1.5 times due to oxidation, while analysis revealed that the unprotected ML-AC coated lines did not change in thickness (**Extended Data Figure 5**). Despite the sharp angle between the bottom surface and sidewall, which can lead to discontinuity due to the lattice-limited curvature[33], ML-AC forms a continuous film that is impermeable to water and oxygen molecules. We reach a similar conclusion for ML-AC on Cu lines by testing chemical stability against copper etchant (**Extended Data Fig. 5**). Next, electrical measurements were performed to check for damage of encapsulated Cu lines during growth. Instead, ML-AC encapsulation improved line conductivity by 25% and current density by 22%, indicating that the low temperature growth process makes ML-AC compatible as a barrier or capping layer material for metal lines (**Extended Data Fig. 5**).

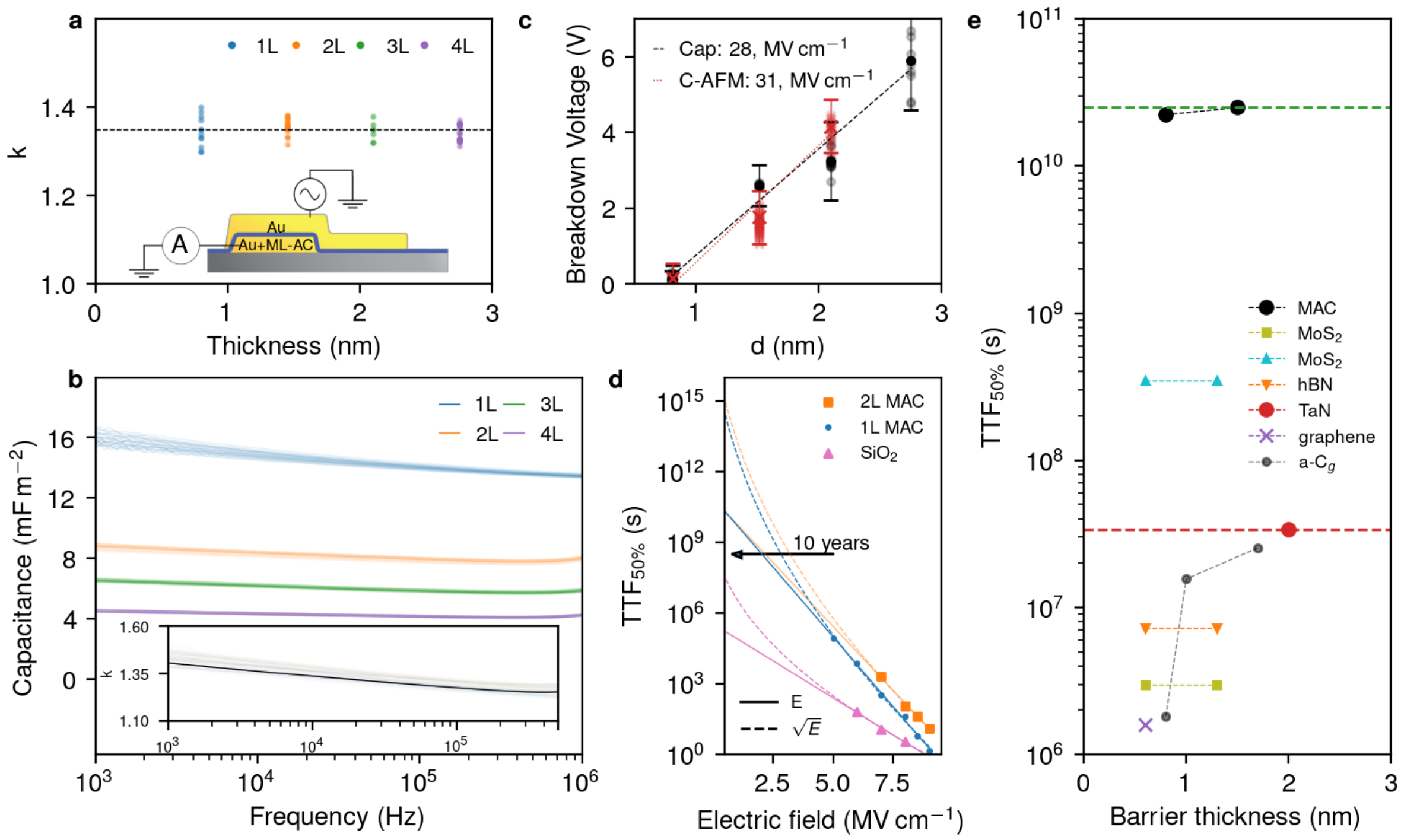

**Figure 4. Dielectric and diffusion barrier performance of directly grown ML-AC. a,** Low-frequency dielectric spectroscopy data for relative dielectric permittivity (at 10 kHz) as a function of thickness (dashed line indicates average value of 1.35). Inset: Schematic of MIM device used. **b,** Frequency dependence of the capacitance normalised per plate area. Inset: relative dielectric permittivity calculated from these capacitances. **c,** Breakdown voltage dependence on ML-AC thickness for measured by MIM devices (black dots and dashed line) and conductive probe AFM (red crosses and dotted line), which gives dielectric strength of 28 and 31 MV cm$^{-1}$ respectively. **d,** Lifetime prediction from copper ion diffusion experiment shows time-to-failure (TTF) improvement by two orders of magnitude over existing 10 years benchmark when ln(TTF) ~ E and ln(TTF) ~$E^{1/2}$ models are applied. For comparison, the substrate ($SiO_2$) behaviour is illustrated by the purple line. **e,** Comparison of the linear model fit of TTF for recently reported replacements of conventional diffusion barrier (transferred graphene (our data), TaN[34], grafted amorphous carbon[34], transferred $MoS_2$[35], transferred hBN[36]).

Finally, the ultralow-k properties of ML-AC were investigated using impedance spectroscopy with metal-insulator-metal (MIM) capacitor devices (**Fig. 4a**), featuring contact areas ranging from 0.1x0.1 μm$^2$ to 50x50 μm$^2$. The thicknesses of the ML-AC layers in the measured MIM devices were determined by cross-sectional TEM (**Extended Data Fig. 6**), and these values agreed with AFM and ellipsometry measurements. The capacitance of ML-AC remains independent of the applied voltage within the ±1 V range at frequencies of 1 kHz, 100 kHz and 1 MHz (**Extended Data Fig. 6**). The dielectric constant of ML-AC was determined by directly measuring the capacitance in the low-frequency domain (1 kHz to 1 MHz) using an AC bias of 0.3 V with an applied DC bias of 1 V. The average dielectric constant was found to be 1.35 (± 0.10 for 4 layers) (**Fig. 4a**), independent of the layer number and with minimal frequency dependence (**Fig. 4b**). This thickness independence contrasts with observations in 3D amorphous thin films or 2D layered crystals. For single-layer MAC, the intrinsically high direct tunnelling current associated with sub-nm thicknesses can influence direct capacitance measurements. Therefore, to verify the dielectric constant, we used a Au/0.8nm MAC/4.8nm $SiO_2$/Si device structure. Fitting the impedance hodograph (5MHz to 1Hz) to an equivalent L-R/C circuit model yielded a dielectric constant of 1.35 ± 0.34, consistent with values obtained from MIM devices (**Extended Data Fig. 6**). This unique ULK behaviour, even for a single layer, can be attributed to the intrinsic low polarizability and low density of MAC. Notably, the dielectric constant of ML-AC (1.35), is considerably lower than that of 2D crystalline hexagonal boron nitride (hBN, 3.3) and the recently reported 3D a-BN (1.78). To confirm that the ULK value originates from the intrinsic properties of individual MAC layers and not from interlayer bonding, we prepared and measured layer-by-layer transferred, 4-layer MAC[24] samples. The dielectric constant of this manually stacked MAC is comparable to that of directly grown ML-AC, supporting this conclusion. (**Extended Data Fig. 6)**.

The low-frequency k values of ML-AC align well with the spectroscopic ellipsometry results (**Extended Data Fig. 6)**. The Cody-Lorentz-Urbach model commonly used for amorphous materials[36] was applied to Ψ-Δ spectra collected at 7 different incidence angles. The permittivity spectra reveal that the real part dips below 1.35. The model reports an increase in imaginary and real parts of the relative dielectric permittivity towards the UV region, coinciding with higher absorption in this range[24] (**Extended Data Fig. 6**).

To understand the unique electronic properties of ML-AC, we refer to the NEXAFS data shown in **Fig. 2f**. Compared to graphene, the angular dependence of 1s-π* transition is weak for single-layer ML-AC and negligible for four layers. The suppressed angular dependence indicates that the π and $\sigma$ orbitals are no longer spatially aligned, likely due to out-of-plane distortions and the resulting randomly oriented electron orbitals within the amorphous structure of ML-AC[37]. This random orientation suppresses dipole formation, leading to low polarizability and, consequently, an ultralow dielectric constant. Furthermore, the typically delocalised $sp^2$ π-orbitals, responsible for electrical conductivity in graphene, may become localised due to this loss of alignment, explaining the insulating nature of ML-AC. For the monolayer case, differences may arise from the substrate influence, which merit further investigation in future studies.

As dielectrics becomes thinner, it is not only necessary to suppress parasitic capacitance with ULK, but also to ensure a high dielectric strength to avoid high leakage currents. We employed two methods to investigate the leakage currents and breakdown voltages of ML-AC. Firstly, we performed slow voltage sweeps until an irreversible breakdown occurred in MIM capacitors with varying plate area and ML-AC thickness (**Extended Data Fig. 7**). We observed similar leakage current densities for various capacitor plate areas for each investigated thickness, demonstrating the homogeneity of ML-AC. Following established protocols[9,38], we defined the breakdown voltage at which a device undergoes an irreversible resistance change, either dropping to very low values or increasing sharply.

Secondly, we used conductive probe AFM (C-AFM) on ML-AC on a gold substrate (**Extended Data Fig. 8**) to eliminate the impact of contamination from device fabrication. At low bias voltage, where the tunnelling barrier remains largely unaffected by the applied electric field, the direct tunnelling current exhibits a linear dependence on voltage V. We derived the breakdown voltage using the method proposed by Lee et al.[39] for hBN, where the breakdown voltages were determined at the transition region from linear to exponential in the I-V dependence. Our measurements indicate that this corresponds approximately to a 12 pA current threshold. Compared to the first method, this provides a more conservative estimate for the dielectric strength, as it does not cross the point of irreversible resistance change.[38]

In both measurements, the dielectric strength of ML-AC was determined by linearly fitting the dependence of breakdown voltage on the thickness, thereby eliminating interface contributions. This resulted in a value of 28 MV $cm^{-1}$ and 31 MV $cm^{-1}$ for MIM geometry and C-AFM measurements, respectively (**Fig. 4c**), which are among the highest reported for any material. The breakdown voltages were also presented in a Weibull plot[38], where the slope, $\beta$, was greater than 4 in all cases, indicating low variability and highly predictable results (**Extended Data Fig. 7, 8**). Furthermore, each most probable breakdown voltage, as indicated by the intercept of horizontal dotted line in the Weibull plot, coincides well with the breakdown voltage estimates in **Fig. 4c**.

Interestingly, the breakdown strength of ML-AC even surpasses that of diamond. This is consistent with the general observation that lower dielectric constants correspond to increased breakdown strength[42]. The random arrangement of strong in-plane covalent bonds and weak interlayer interactions in ML-AC are likely to impede electron movement within the layers and hinder charge transfer between layers. This unique structure can account for the low polarizability and minimise material ionisation, the primary mechanism for dielectric breakdown.

Finally, we investigate the leakage current behaviour and observe an exponential scaling with thickness, such that each additional MAC layer reduces the current density by approximately 3 orders of magnitude. At 0.5 V, 1-layer shows a current density of 10 A $cm^{-2}$, 2-layer shows $10^{-2}$ A $cm^{-2}$, 3-layer shows $10^{-5}$ A $cm^{-2}$, and 4-layer extrapolates to $10^{-8}$ A $cm^{-2}$, respectively (**Extended Data Fig. 7**). With this, we conclude that the leakage current for ML-AC is dominated by direct tunnelling. We utilised the low power limit for transistor operation, where current densities under $1.5x10^{-2}$ A $cm^{-2}$ would meet the off-current threshold with at least 0.7 V operating voltage[15,40,41]. Therefore, we demonstrate that 2-layer (1.4 nm) and thicker ML-AC in Au/ML-AC/Au capacitors are already within the leakage current limit for transistor operation. Interestingly, the leakage current of ML-AC surpasses that of exfoliated hBN at an equivalent thickness, aligning closely to the theoretical value of hBN[41].

The hardness of ML-AC was measured using AFM in nanoindentation mode. The force displacement curves were obtained by indenting on $SiO_2$ and ML-AC on $SiO_2$ with a diamond tip. The hardness of ML-AC was found to be at least one order of magnitude greater than that of $SiO_2$, approximating to ~100 GPa. This, coupled with the previously reported Young's modulus of 200 GPa[24], distinguishes ML-AC from the other potential ULK materials. Existing ULK materials are brittle due to porosity, which poses a significant challenge for scalability within industry processes (**Extended Data Fig. 9**).

Interconnect design needs additional functional layers to facilitate metal line formation and prevent metal ion degradation (**Extended Data Fig. 5**). Therefore, we further examined the performance of ML-AC as a metal (Cu) interdiffusion barrier, by statistically analysing current versus time at various applied bias voltages (**Fig. 4d** and **Extended Data Fig. 10**). Time-to-failure (TTF), defined by probability of 0.5 and determined by a linear ln(TTF) ~ E model, indicated failure times of $10^{10}$ s at operating electric fields of 0.5 MV $cm^{-1}$. ML-AC meets the requirements for metal ion diffusion barrier and outperforms all recently reported and commonly used materials by at least two orders of magnitude[35,43,34] (**Fig. 4e**). The potential implementation of ML-AC as an inter-metal dielectric in interconnects could be realised with a new architecture that eliminates the need for functional layers at the ULK-to-metal interface and reduces the dielectric spacer volume to enable the packing of a higher density of transistors. This also maximises the available volume for Cu or Co lines to meet line conductivity requirements for improved performance by reducing RC delay and heat generation.

In conclusion, we have shown that ML-AC is a promising 2D ULK dielectric material with significant potential in the semiconductor industry. Industry is already considering numerous uses of 2D materials, and ML-AC stands out due to its exceptionally low relative dielectric permittivity, high dielectric strength, and excellent mechanical robustness. The integration of ML-AC is made possible by its technologically relevant characteristics, such as precise control over layer-by-layer thickness, scalability down to sub-nanometre dimensions with thickness independent dielectric constant, and the ability to grow directly on dielectric surfaces at low temperatures in a conformal manner. While ML-AC represents a significant milestone for use as a low k dielectric in microelectronics, it also opens new possibilities for future developments in 2D electronics.

## Acknowledgements

B.Ö. acknowledges support by the National Research Foundation, Prime Minister's Office, Singapore, under its Competitive Research Programme (CRP award number NRF-CRP22-2019-008) and Medium-Sized Centre Programme (CA2DM), by EDB Singapore, under its Space Technology Development Programme (S22-19013-STDP), and by the Ministry of Education of Singapore, under its Research Centre of Excellence award to the Institute for Functional Intelligent Materials (I-FIM, Project No. EDUNC-33-18-279-V12). N.K., Y.S. and K.S. acknowledge support by JST CREST (JPMJCR20B1). M.B. acknowledges support by the Ministry of Education of Singapore, via Academic Research Fund Tier 2 (MOE-T2EP50122-0016).

## Author contribution

B.Ö. initiated the work. B.Ö. and C.-T.T. coordinated the work. C.-T.T. and A.K.G. contributed equally to the work. C.-T.T., A.K.G and B.Ö. designed the experiments. A.K.G. performed ellipsometry, impedance spectroscopy and dielectric strength determination experiments and all device fabrication. U.K. performed Raman spectroscopy, copper-diffusion determination experiments and sample growth optimization. A.K.G., U.K., H.Z., A.A., L.S., C.C.T, and A.S. synthesised the material. U.B, H.Y. and M.B. performed cross-sectional TEM and EELS measurements. N.K., Y.S. and K.S. performed atomic resolution STEM imaging. D.V.V., A.M., A.F. performed synchrotron-based XPS and NEXAFS measurements at BESSY-II. C.-T.T., A.K.G. and B.Ö. analysed all data and wrote the manuscript. All authors discussed and commented on the manuscript.

# Materials and Methods

## Synthesis of ML-AC

The CVD setup uses an excimer laser (XeCl, λ = 308 nm, 50Hz pulse rate) and remote inductively coupled plasma source for synthesis. Gases were introduced with a ratio of $C_2H_2$:Ar at 1:2 to maintain a pressure in the range of $10^{-3}$ mbar. Plasma was tuned to the lowest operating power that will maintain a discharge glow. Laser was attenuated to maintain substrate surface temperature within 250°C - 300°C in cases for direct laser exposure to the growth substrate. Laser fluence of 300 mJ $cm^{-2}$ and stage heating to 250°C - 300°C was used in cases for laser directed away from the growth substrate. Direct laser exposure was typically used for flat Cu foils, Si and Si/$SiO_2$ substrates while indirect laser exposure was used for lithographically patterned devices and large area growths (>1 cm). Both cases yield similar ML-AC structure.

The synthesis of single-layer MAC followed the growth method detailed in reference 24. Subsequent transfer process involved the following steps. MAC was spin-coated with PMMA (495 PMMA, A4, 4% in anisole) at 4,000 rpm to serve as a polymer support during the transfer process. The spin-coated samples were then baked on a hotplate at 180 °C for 2 mins. MAC from the opposite side of the substrate was removed by 50 W Ar plasma for 5 mins. The Cu foil was wet etched in a 0.7 wt% ammonium persulfate solution $(NH_4)_2S_2O_8$, by floating the growth substrate on the solution with MAC facing up. The MAC was transferred with a rigid substrate to deionized water for two 2 h rinses before scooping onto the desired substrate ($SiO_2$/Si wafers or holey SiN TEM grids). The PMMA support was removed by submerging the sample in acetone for 30 mins. This transfer process was repeated to manually stack additional layers of MAC. The samples were annealed in 5% $H_2$ in Ar gas for 5 hours to remove any organic residues.

## Transmission electron microscopy characterisation

Cross-sectional TEM imaging samples were prepared by two approaches, either by hand polishing followed by Argon ion milling at room temperature, or by Helios 450S Focused

Ion Beam with the sample cooled down to liquid nitrogen temperature. Gold (90-150 nm) was deposited by thermal evaporator with a water-cooled stage at a deposition rate of 0.2 A/s to minimise influence on the carbon layer. A thin layer of platinum (Pt) was subsequently deposited on the surface of the sample (Si/SiO2/ML-AC/Au/Pt) to further minimise ion milling damage. The cross-section of the deposited region was then lifted off and mounted onto the TEM grid. Gallium (Ga) ions accelerated at 30 kV at various currents were then used for the initial thinning of this lamella. Low energy cleaning (2 kV) was used at the final stage to remove redeposition and amorphization. The thinnest FIB-cut lamella was achieved for the thickest ML-AC film, providing the best resolution of the layers.

A ThermoFisher Titan TEM/STEM with Schottky electron emitter was operated at 300 kV for imaging with a Gatan OneView camera, and EELS was performed with a Gatan Tridiem spectrometer using 10 mrad STEM convergence semiangle and an on-axis 15 mrad EELS collection semiangle. STEM images were taken before and after EELS, to examine possible beam damage or carbon contamination and to check for sample drift during the EELS measurement. To reduce carbon contamination before thickness determination via EELS, cross-sectional samples were cleaned in a 25% $O_2$: 75% Ar plasma for several minutes prior to STEM measurements.

Atomic resolution STEM samples were transferred from corresponding substrates to silicon nitride grids and annealed in $Ar/H_2$ atmosphere at 0.3 mbar and 300 °C for 4 hours. For the case of Cu, ammonium persulfate was used to etch away the substrate. For the case of $Si/SiO_2$, a concentrated KOH solution was used to etch the silicon dioxide layer. The atomic structure of 1L, 2L and 3L ML-AC was observed using an ultra-high vacuum ARM200F-based microscope equipped with a JEOL delta aberration corrector and a cold field emission gun (TripleC#3 microscope, JEOL). The acceleration voltage was set to 60 kV, and the specimen was heated to 500°C using a heating holder during STEM observation.

## Structural characterisation

Bruker Dimension Icon, Bruker Dimension Fastscan and Park XE-120 atomic force microscopes were used to investigate our samples. Thickness was measured with Fastscan operated in tapping mode with Bruker FastScan A probes. Morphology was measured with Bruker Icon in PeakForce tapping mode using Budget probes Multi75AlG and Park XE-120 in True Non-Contact mode. Measurements of samples with varying growth time were performed with 512-pixel resolution and averaging of the profile with 48-pixel wide line cut. Conductive AFM was measured with PeakForce TUNA module and Pt covered tips: OSCM-PIT (15 nm tip radius) and SCT-PIT-V2 (25 nm tip radius) tips by Bruker. Additionally, measurements were reproduced using RockyMountain full-metal body tips with a 10 nm tip radius. For each measurement, a pure metal surface was used to check that tip conductivity remained unchanged.

Raman spectra were acquired with a WITEC Alpha 300R using an excitation of a 532 nm laser and a 1µm diameter circular laser spot at room temperature. Measurements were carried out using a laser power of ~1.1 mW and a total accumulation time of 30 s.

The X-ray photoelectron spectroscopy (XPS) experiment was carried out using a Thermo Fisher Scientific K-Alpha instrument (USA) and Kratos DLD Ultra XPS. The analysis chamber vacuum was maintained at $5 \times 10^{-10}$ Pa. The excitation source used was Al Kα radiation (hv=1486.68 eV), with a working voltage of 15 kV and a filament current of 10 mA. Signal accumulation was performed over 5-10 cycles. The pass energy was set to 50 eV with a step size of 0.05 eV. The binding energy was calibrated by using substrate peaks, such as Si 2p (103.5 eV), Cu 2p (932.6 eV) and Co 2p (778.2 eV) to restore the position of C 1s lines due to the insulating nature of ML-AC and Si/$SiO_2$ wafers and subsequent charging.

Room temperature NEXAFS measurements were performed at the Russian-German dipole beamline of the BESSY-II electron storage ring operated by Helmholtz–Zentrum Berlin für Materialien und Energie.[44] NEXAFS spectra were recorded in total electron yield mode. The measurements were performed for the annealed (~500 °C) samples of ML-AC on $SiO_2$. All the spectra were pre- and post-edge normalised.

## Nano- and micro-device fabrication

JEOL Electron-beam lithographer JBX-6300FS was used to pattern trenches, cobalt and copper lines, and devices mentioned in the text. For trenches and metal lines patterns, CSAR 6200 resist was exposed to 500 µC/cm$^2$ at 100 keV. For MIM devices, bilayer PMMA A5 495k and PMMA A5 950K e-beam lithography resists were exposed to 1650 uC/cm$^2$ at 100 keV. Bottom electrodes were deposited using AJA e-beam evaporator at $10^{-7}$ mbar conditions for 2nm Ti (at 0.2 A/s) and 58 nm Au (at 0.3 A/s). Lift-off was performed in acetone followed by chloromethane. Substrate was cleaned with $O_2$ plasma (20 W, 20 sccm, 40 sec). ML-AC growth was performed, followed by another step of bilayer resist lithography to define the top electrodes. Metal deposition was repeated under the same conditions but without Ti layer. For the case of metal lines, Co (at 0.2 A/s), Cr (at 0.2 A/s) and Cu (at 0.6 A/s) were also deposited using AJA e-beam evaporator. Oxford Deep RIE station was used for $CHF_3/SF_6/Ar$ plasma etching of trenches in $SiO_2$.

## Dielectric spectroscopy

For all dielectric constant and dielectric strength measurements, except for the dielectric constant of MAC measured using the Au/MAC/$SiO_2$/Si structure, we utilized 1L to 4L ML-AC sandwiched between Au electrodes. The thickness of MAC in Au/MAC/$SiO_2$/Si devices was determined from 10px averaged cross-sectional TEM EELS line profile at the FWHM. Similarly, the thickness of ML-AC in all other Au/ML-AC/Au MIM devices was obtained from 50px averaged cross-sectional TEM line profiles at the FWHM. The dielectric constant of ML-AC was calculated using the measured capacitance and the thickness, applying the equation: $\varepsilon = \frac{Cd}{A\varepsilon_0}$, where $C$ is the capacitance, $d$ is the thickness, $A$ is the electrode area, and $\varepsilon_0$ is the permittivity of free space.

A Zurich Instruments MFIA impedance analyser with $10^3$ V/A - $10^{12}$ V/A transimpedance converters were used to measure highly resistive scaling capacitor junctions of Co-(ML-AC)-Co, Au-(ML-AC)-Au and Si/$SiO_2$-(ML-AC)-Au. A frequency range of 1 kHz - 1 MHz was used since <1 kHz data were too noisy and >1 MHz results had additional contributions from the connections and measurement setup that interfered with the data. A simple L-(R/C) model, which corresponds to a capacitive behaviour of the formed structure was used for analysis. We took into account the presence of any oxide layers formed at the metal interface to ML-AC to exclude their contribution to the measured dielectric permittivity.

An Accurion ellipsometer was used to track the dielectric permittivity behaviour in the visible range. Ψ-Δ spectra at 7 different incidence angles were collected and fitted using Cody-Lorentz-Urbach (CLU) model. Thickness was fixed according to cross-sectional TEM measurements and band gap energy at ~2.5 eV according to our UV-vis absorption measurements. Other parameters for CLU were: A = 14.08 eV, $E_0$ = 5.269 eV, Gamma = 7.85 eV, $E_p$ = 0.458 eV, $E_t$ = 0.714 eV, $E_u$ =1.718 eV.

## Cu diffusion barrier

Copper diffusion devices with 200 um diameter circular contacts were fabricated on heavily doped 90 nm $SiO_2$/Si using metal masks with a Lesker Nano 36 Thermal Evaporator (Device: Al (20 nm)/Cu (80 nm)/ML-AC/$SiO_2$/Si ). *I-V* and *I-T* characteristics were measured in ambient condition using a probe station with a Keithley 6430 Sub-femtoamp Remote Sourcemeter.

# Extended Data

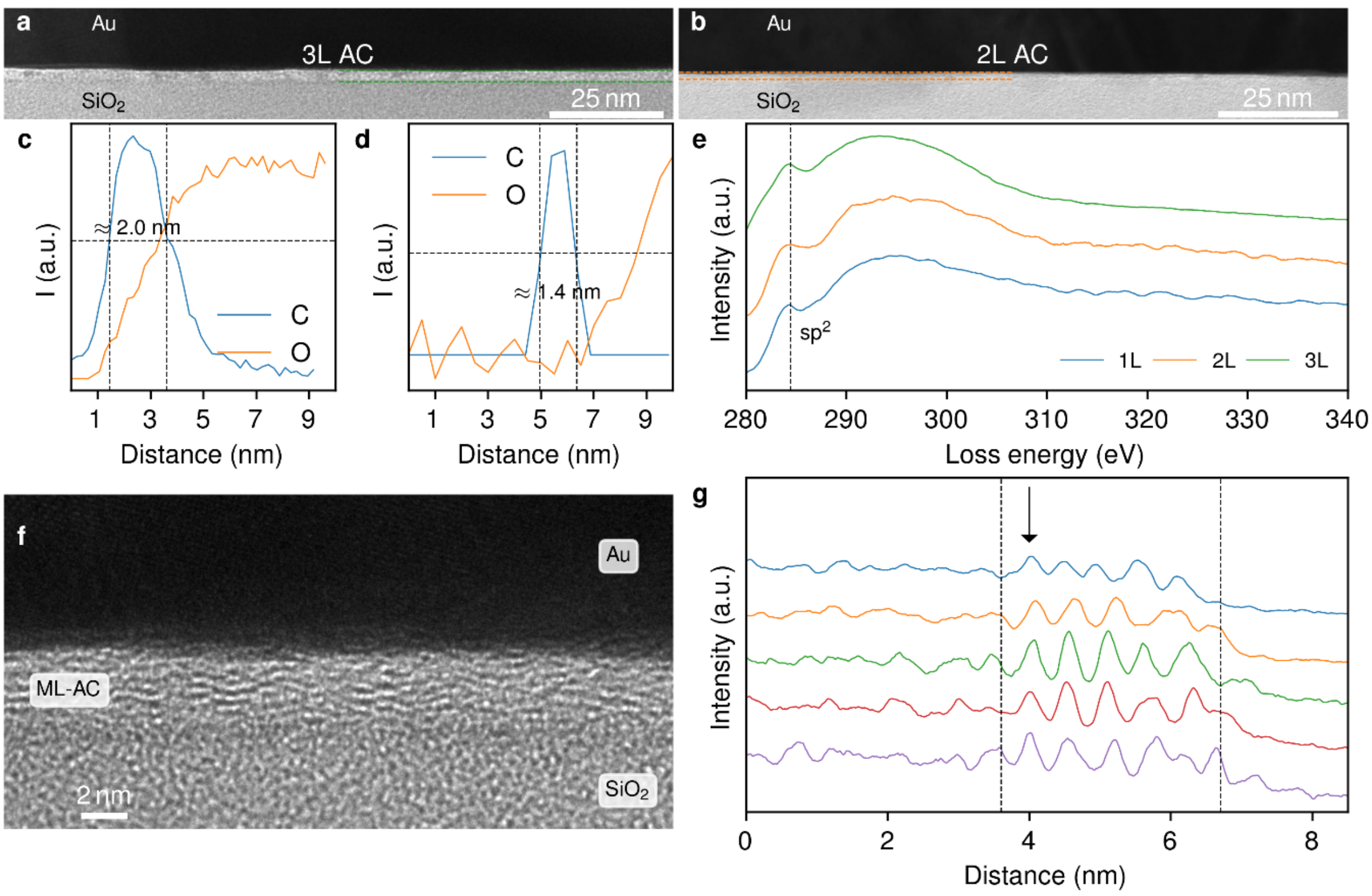

**Extended Data Figure 1**. **Cross-sectional TEM analysis of ML-AC. a-b,** Cross-sectional TEM images of 3- and 2-layers ML-AC on $SiO_2$/Si wafer with **c-d,** corresponding Carbon and Oxygen K-edge EELS intensity line profiles, respectively. ML-AC has uniform thickness for over >100 nm field of view. **e,** EELS carbon K-edge spectra corresponding to **Fig. 1c-e,** with black dotted line indicating peak position for $sp^2$ bonds. **f,** Cross-sectional HRTEM image of 5-layer ML-AC on $SiO_2$/Si wafer and **g,** corresponding HRTEM intensity line profiles averaged over 10 pixels obtained from **f** (first layer of ML-AC is indicated by the arrow).

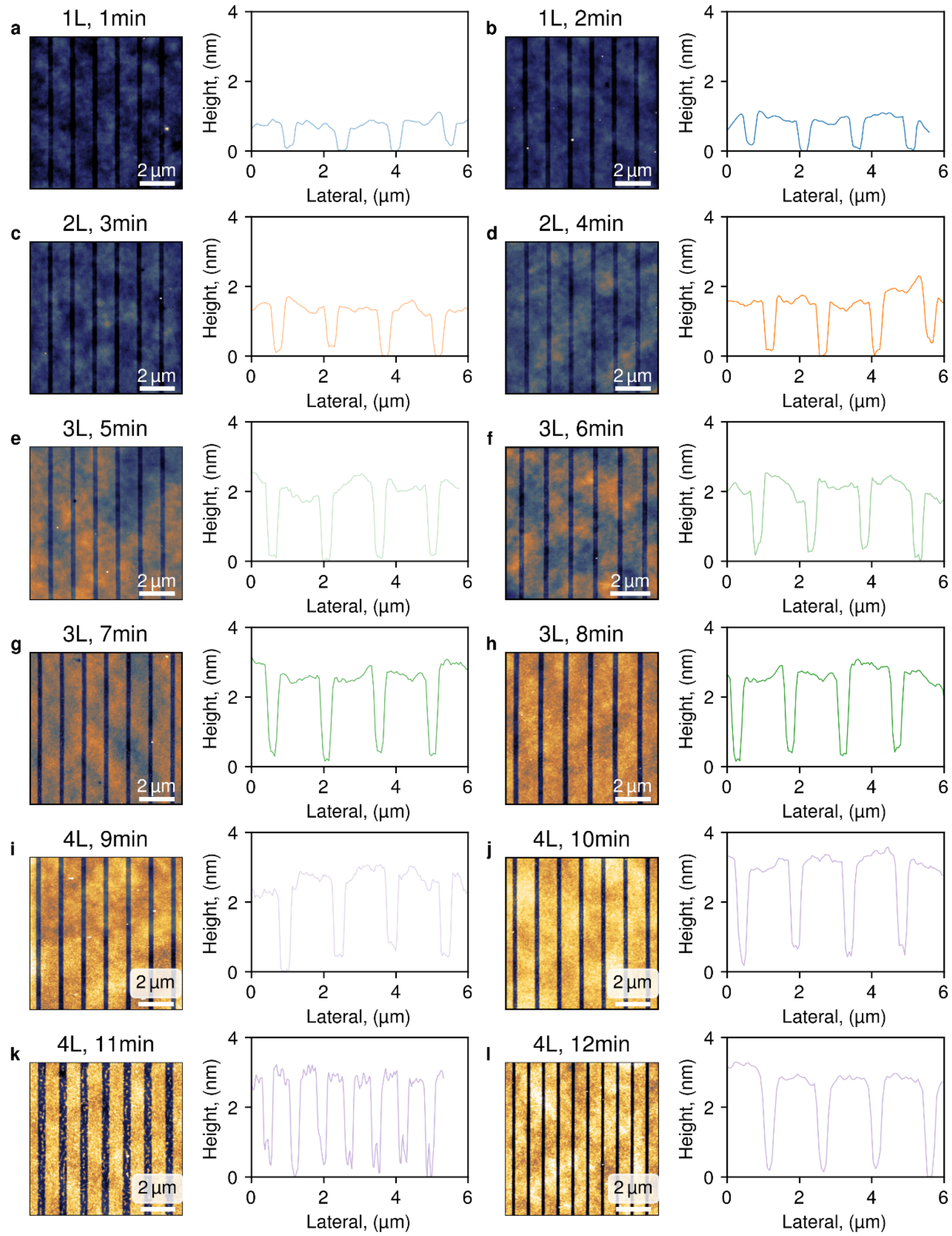

**Extended Data Figure 2. Thickness measurements of ML-AC. a-l,** AFM topography images and corresponding profiles for 1 to 12 min of ML-AC growth on $SiO_2$/Si substrates with 1-minute intervals. Etched regions were defined by electron beam lithography and oxygen reactive ion etching.

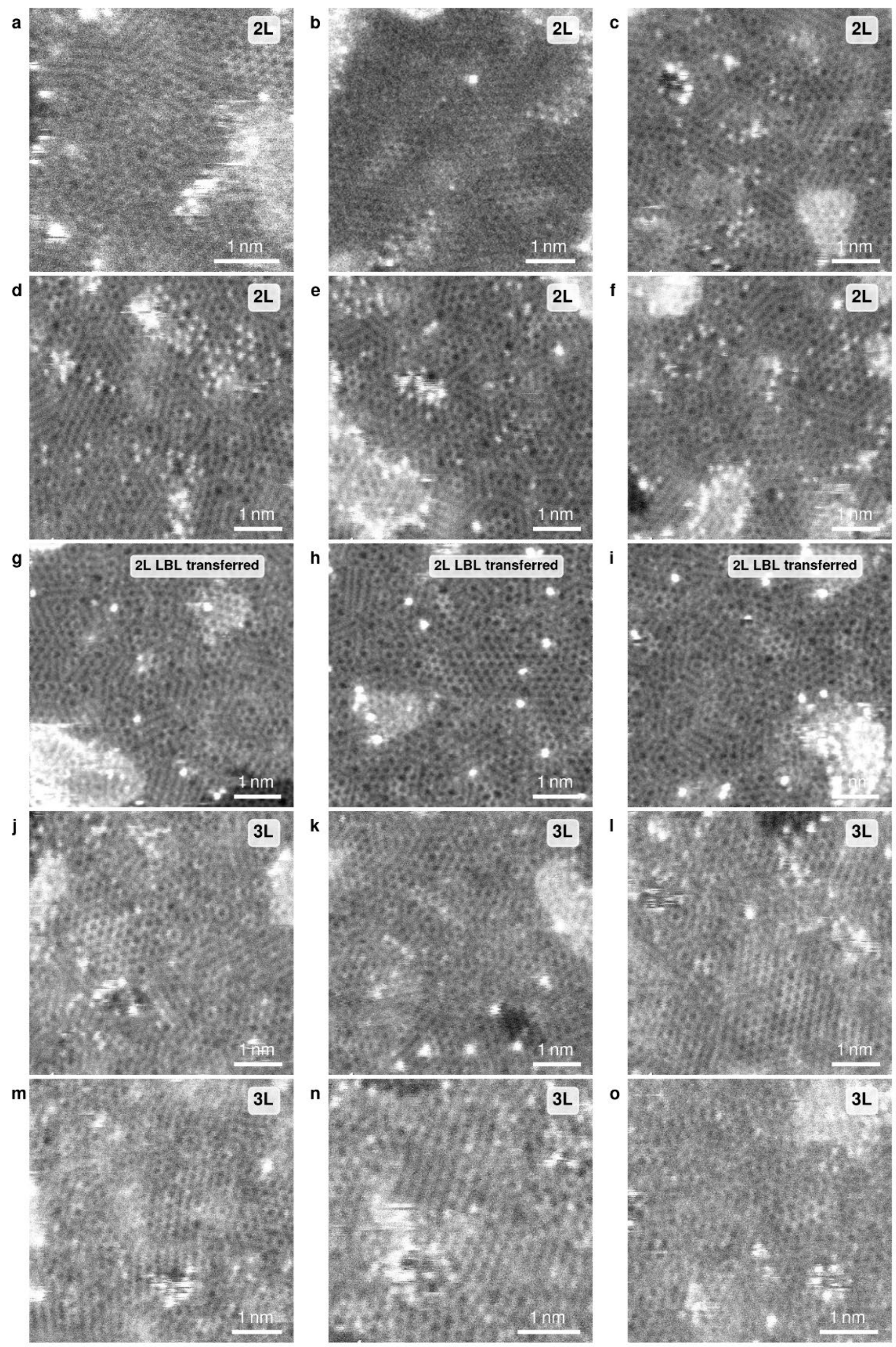

**Extended Data Figure 3**. **STEM imaging of directly grown ML-AC on $SiO_2$ and layer-by-layer transferred MAC transferred to TEM grid.** Atomic resolution images showing moiré interference from **a-f,** as-grown 2-layer ML-AC, **g-i** layer-by-layer transferred 2-layer MAC and j-**o,** as-grown 3-layer ML-AC.

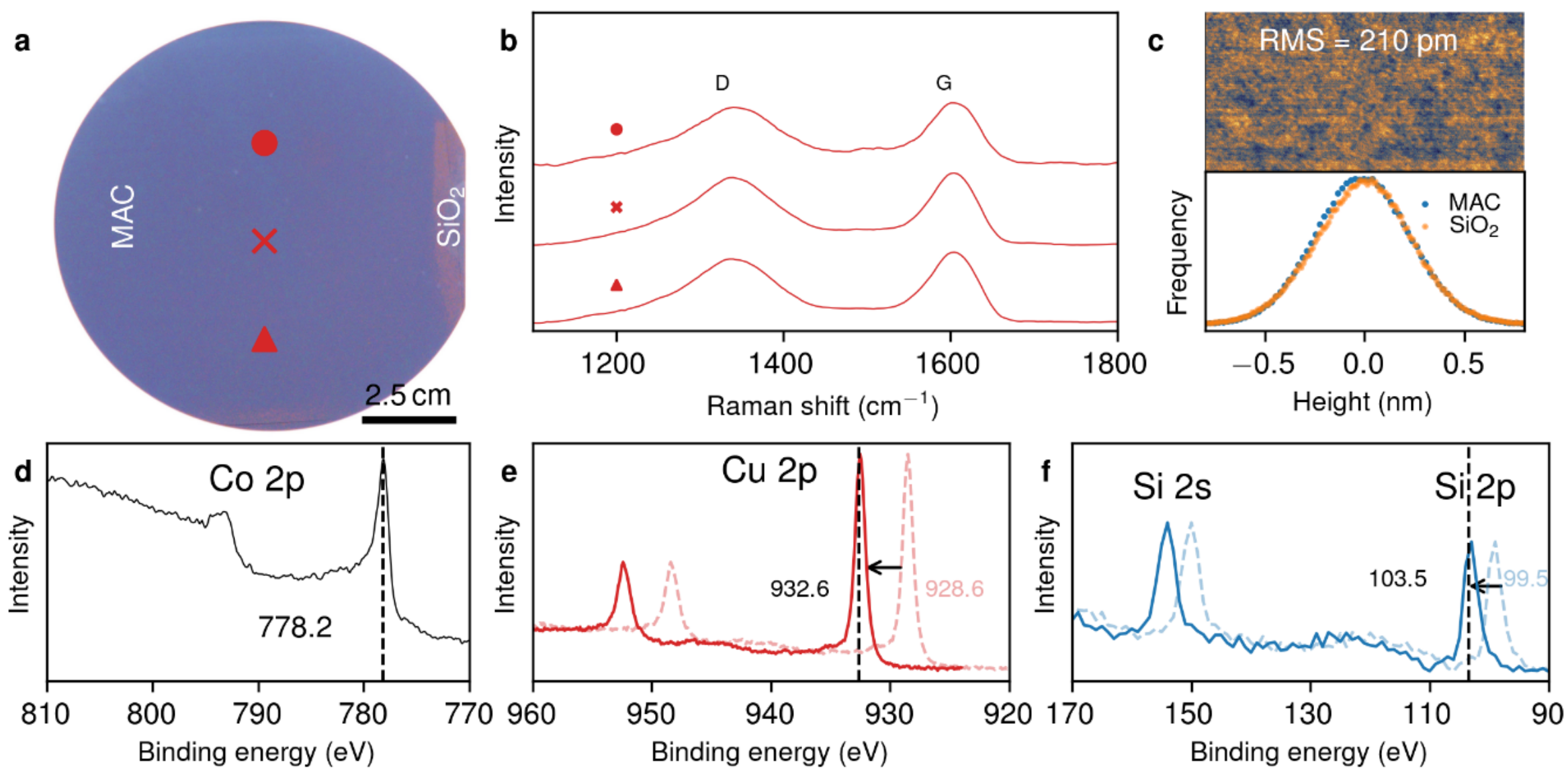

**Extended Data Figure 4**. **Additional spectroscopic data from ML-AC. a,** Optical image of the 2-layer ML-AC grown on 4-inch wafer and **b,** Raman spectra of 3 representative points from corresponding regions indicated by symbols in **a**. **c,** AFM image of the surface of ML-AC grown on $SiO_2$ and the histogram of the height distributions (surface roughness) for $SiO_2$ and ML-AC on $SiO_2$[30]. **d-f,** Core level XPS spectra of the substrate used for calibration of the C 1s line position by shifting the original spectra (dashed lines) to the reference peaks labelled, and to show the absence of bond formation between ML-AC layer and the substrate (by line shape and the spacing between main line and satellite feature).

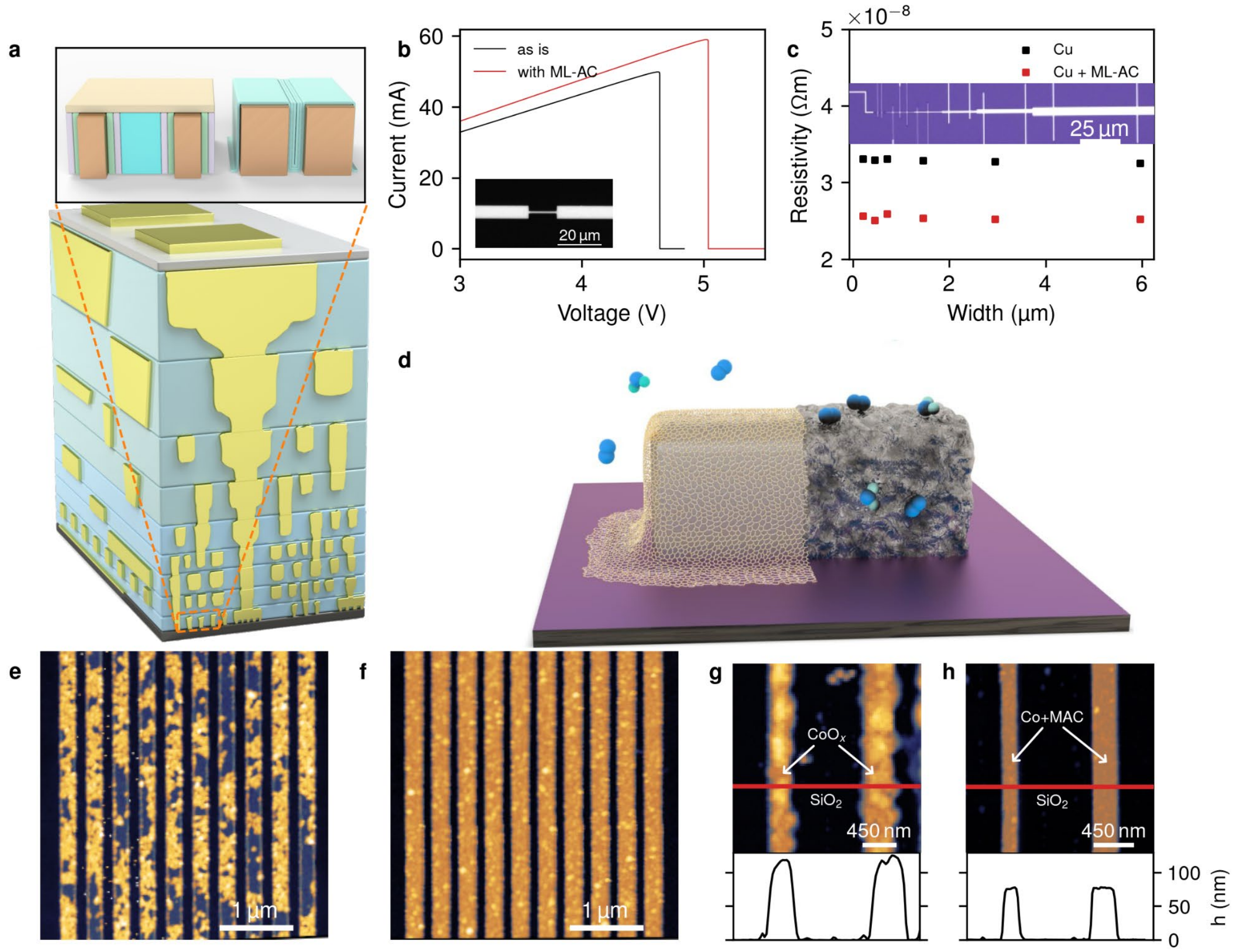

**Extended Data Figure 5**. **Chemical protection of metal lines by ML-AC. a,** Schematic of metal interconnects, with blue colour corresponding to the ultralow-k dielectric surrounding conductive elements. Inset: (left) Complex structure of multiple functional layers surrounding each line is shown. (Right) A proposed simplified structure, where a ML-AC dielectric combines all required functionalities. **b,** I-V measurements show maximum current before breakdown and **c**, Cu resistivity at different line widths for pristine Cu line (black) and Cu line protected by ML-AC (red). **d,** Schematic render of metallic line degradation under ambient conditions. AFM images of Cu lines **e**, without and **f**, with directly grown ML-AC encapsulation exposed to ammonium persulfate etchant has average height of 8 nm and 30 nm, and roughness of 6.5 nm and 1.7 nm, respectively. Height and roughness of ML-AC encapsulated Cu lines remain unchanged before and after exposure to Cu etchant. AFM images of 80 nm thick cobalt lines on $SiO_2$/Si substrate **g**, without and **h**, with 2-layer ML-AC after exposure to ambient conditions for 72 hours.

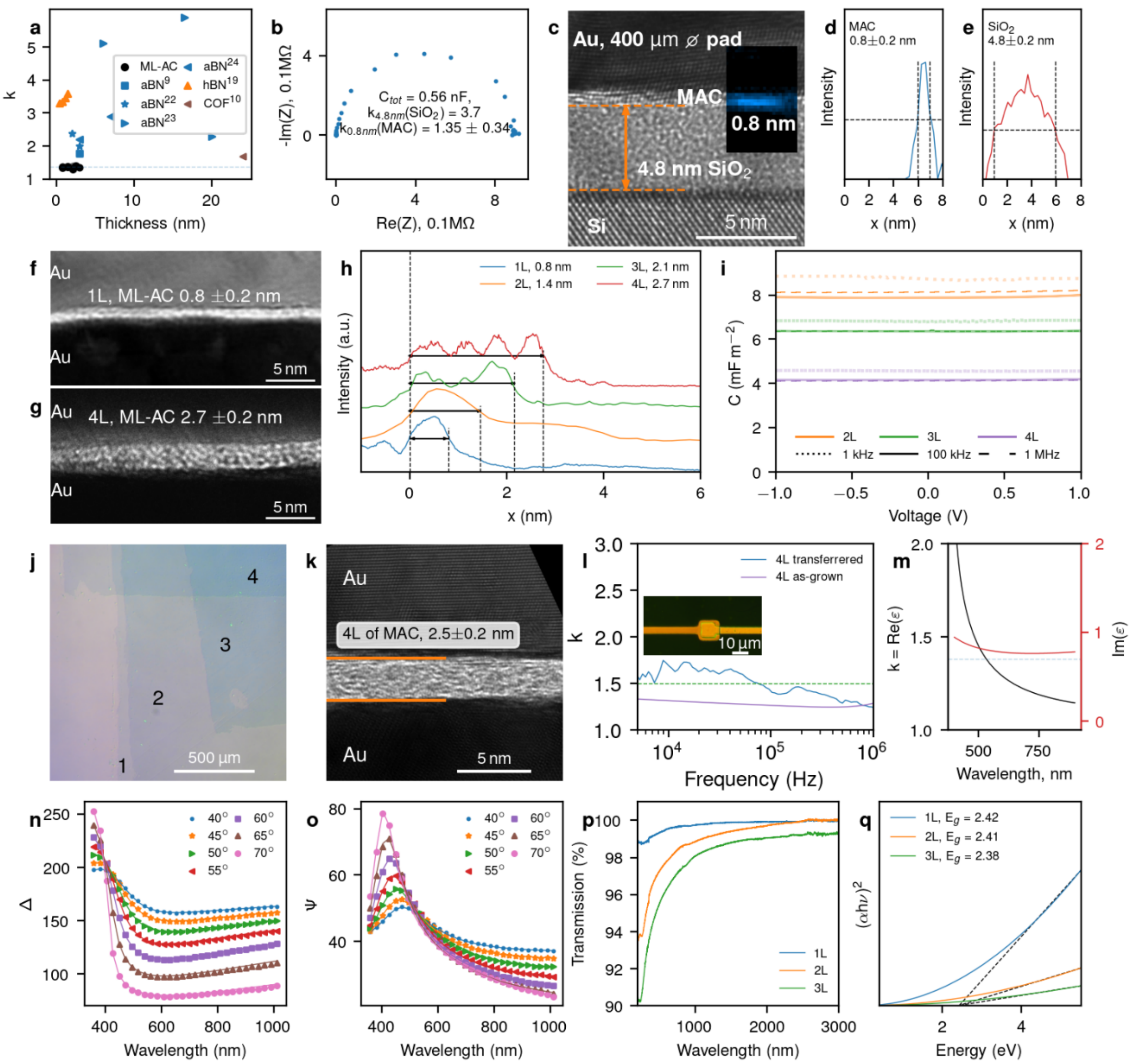

**Extended Data Figure 6**. **Additional data on dielectric spectroscopy. a,** Dielectric permittivity of ML-AC for different dielectric thickness and comparison to hBN, a-BN and covalent organic frameworks. Dashed line corresponds to 1.35. **b,** Impedance hodograph of MAC measured with **c,** Au/MAC/$SiO_2$/Si 400µm diameter device shown in cross-sectional TEM. Overlay: EELS map for carbon. EELS line profiles for **d,** carbon and **e,** oxygen, where thickness of MAC and oxide layer were respectively measured at the FWHM (dashed line). Cross-section TEM images of MIM capacitors with **f,** 1-layer and **g,** 4-layer ML-AC. **h,** Line profiles from cross-sectional TEM images for 1, 2, 3, and 4-layer ML-AC MIM capacitors used for electrical measurements. **i,** C-V measurements show no capacitance dependence on voltage for 2, 3, and 4-layer ML-AC captured at 1kHz, 100kHz and 1MHz. **j,** Optical image of MAC transferred layer-by-layer to form a 4-layers. **k,** Cross-sectional TEM image of the MIM capacitor with layer-by-layer transferred MAC. **l,** Impedance spectroscopy data comparing the dielectric permittivity spectrum for 4-layers of MAC transferred layer-by-layer and as-grown 4-layer ML-AC. Inset: optical image of a typical MIM device. **m,** Spectroscopic ellipsometry data of imaginary and real parts of the relative dielectric permittivity as a function of wavelength. Dotted blue line corresponds to 1.35. Spectra (dots) of the **n**, Ψ and **o**, Δ fitted by Cody-Lorentz-Urbach model (lines). **p,** UV-vis spectra of films transferred to $CaF_2$ substrates, and **q,** corresponding Tauc plots used to extract optical band gap values.

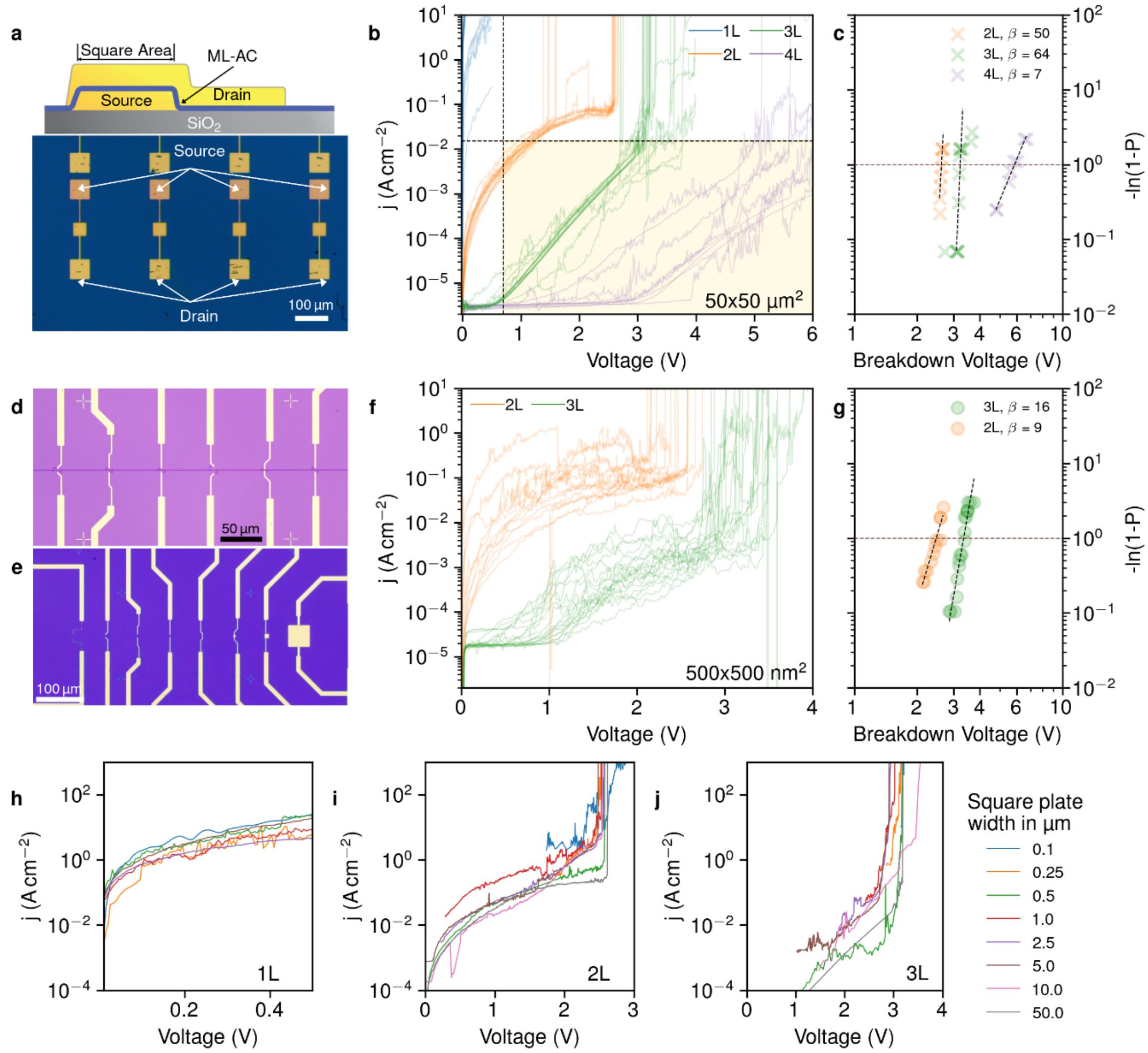

**Extended Data Figure 7**. **Breakdown and leakage of ML-AC measured in MIM devices.** Optical image showing MIM capacitors (Au metal plates) with **a**, 50x50 µm$^2$ **d**, 500x500 nm$^2$ and **h**, varying capacitor plate area. **b**, and **f**, I-V curves measured for devices shown in (**a**) and (**d**) respectively. Horizontal and vertical dashed line in (**b**) represents the low-power limit current density of 1.5 10-2 A cm-2 and 0.7 V transistor operation voltage. Acceptable transistor operational range is within the yellow shaded area. **c**, and **g**, Weibull plots of the breakdown voltage extracted from (**b**) and (**f**) respectively. **h-j,** I-V curves for scaling capacitor plate size for 1, 2 and 3 layers respectively. These data show that leakage current and breakdown voltage does not depend significantly on the plate area.

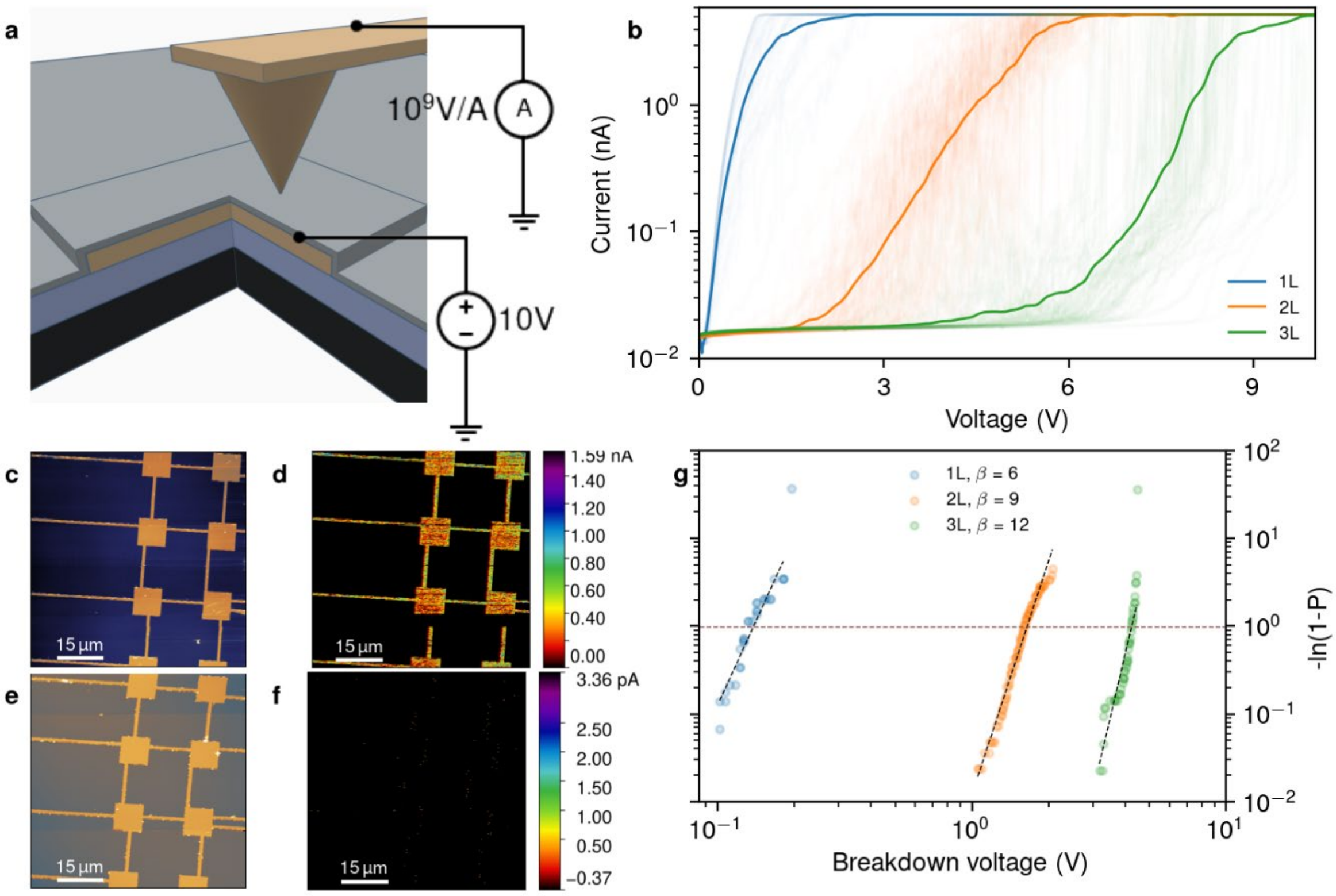

**Extended Data Figure 8**. **Breakdown and leakage current measurement with CP-AFM. a** Schematics of CP-AFM experiment. **b** I-V curves of 1, 2 and 3-layer ML-AC on Au substrate (over 100 points each). Solid lines indicate averaged curves. **c,** Topography and **d,** current map at fixed bias voltage for 1-layer ML-AC. **e,** Topography and **f,** current map at fixed bias voltage for 1-layer ML-AC. Bias voltage for both pairs of images is fixed at 100 mV. **g,** Weibull plot of the data presented in (**b**).

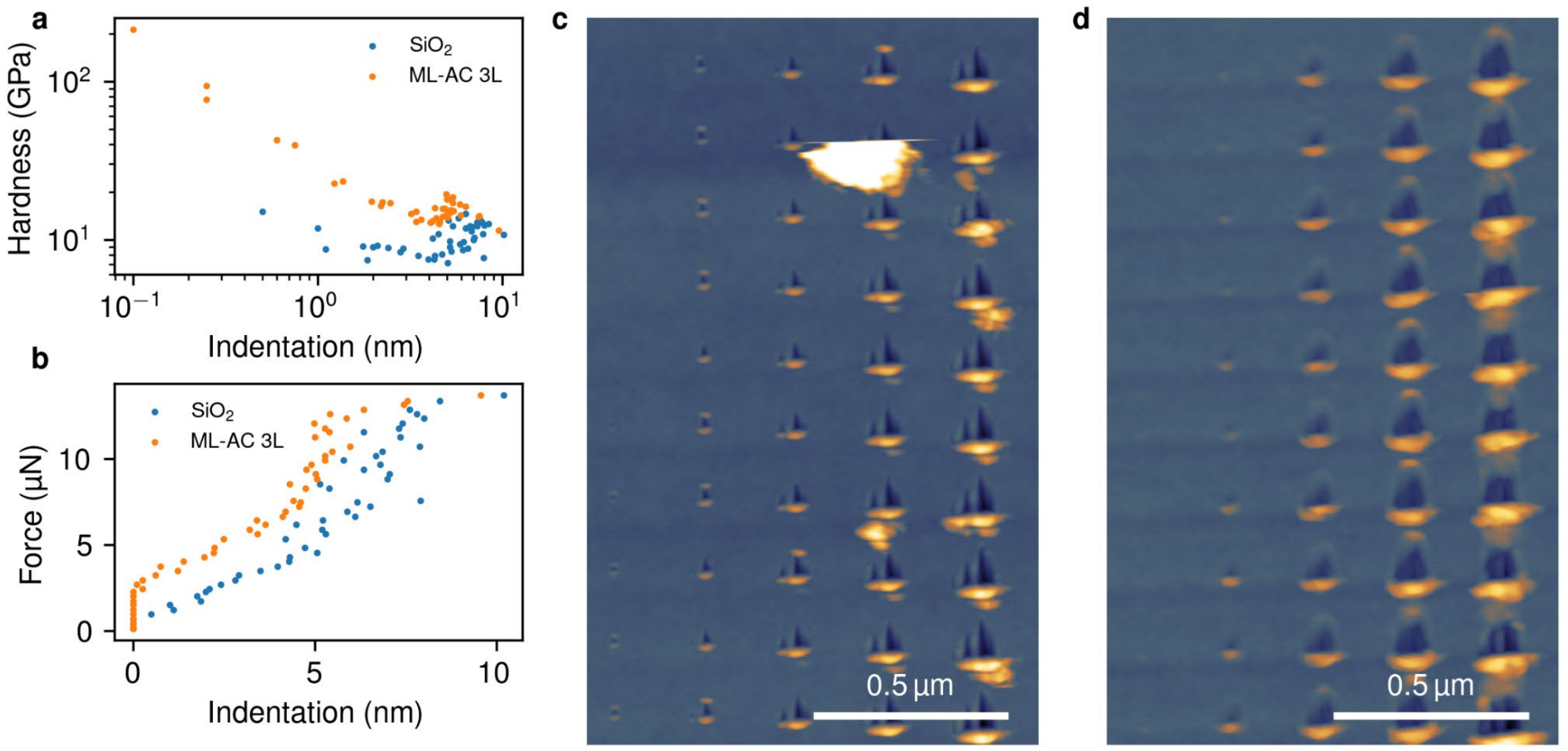

**Extended Data Figure 9**. **AFM based nano-indentation hardness tests. a,** Hardness as a function of indentation depth for 3-layer ML-AC and $SiO_2$ substrate indicating at least one order of magnitude higher values for the case of 3-layer ML-AC **b,** Force-indentation curve and AFM topography images of the indentations formed in **c**, SiO2 and **d**, ML-AC respectively. It is clearly seen that for the ML-AC coated sample, indentations are observed only at much higher force loads.

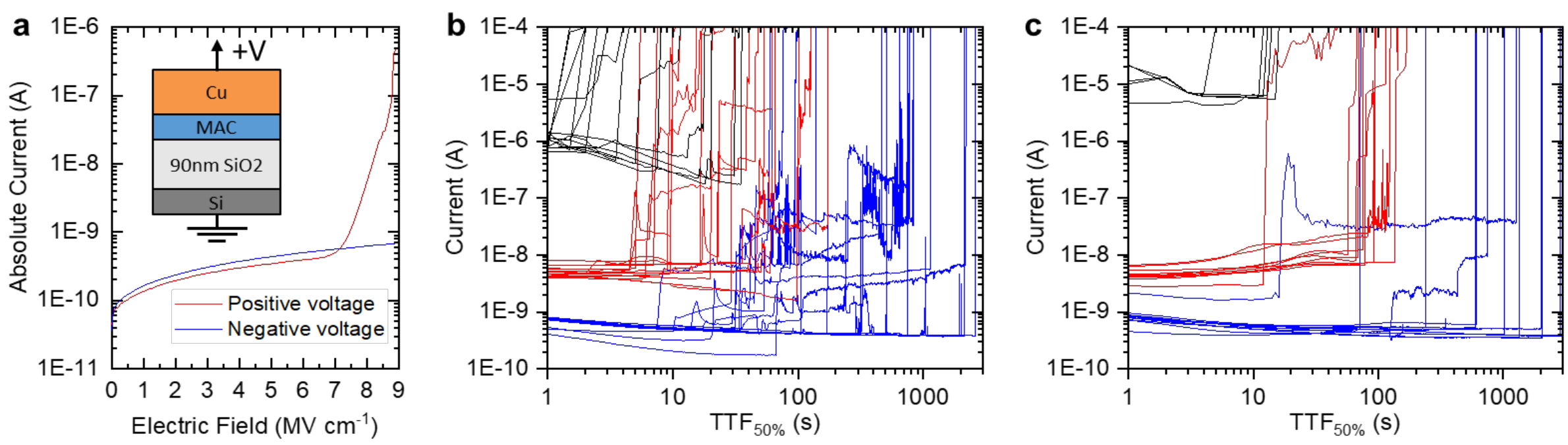

**Extended Data Figure 10. Cu ion diffusion barrier properties of ML-AC directly grown on $SiO_2$. a,** I-V curve of metal-oxide-semiconductor capacitor at 0.5 V $s^{-1}$ sweep rate for both voltage polarity shows Cu diffusion behaviour at positive voltage. 80 nm thick Cu contacts of 200 µm diameter circle are capped with Al (20 nm thick). Inset: Schematic of metal-oxide-semiconductor capacitor for barrier property testing. Current evolution with time for multiple devices with **a,** 1L ML-AC and **c**, 2L ML-AC at applied electric fields of 7, 8 and 9 MV $cm^{-1}$ (Indicated by blue, black and red lines respectively).